# Phonon transport in perovskite $SrTiO_3$ from first principles

Lei Feng, Takuma Shiga, and Junichiro Shiomi*

*Department of Mechanical Engineering, The University of Tokyo, 7-3-1 Hongo, Bunkyo, Tokyo, 113-8656, Japan*

E-mail: shiomi@photon.t.u-tokyo.ac.jp

We investigate phonon transport in perovskite strontium titanate ($SrTiO_3$) which is stable above its phase transition temperature (~105 K) by using first-principles molecular dynamics and anharmonic lattice dynamics. Unlike conventional ground-state-based perturbation methods that give imaginary phonon frequencies, the current calculation reproduces stable phonon dispersion relations observed in experiments. We find the contribution of optical phonons to overall lattice thermal conductivity is larger than 60%, markedly different from the usual picture with dominant contribution from acoustic phonons. The mode- and pseudopotential-dependence analysis suggests the strong attenuation of acoustic phonons transport originated from strong anharmonic coupling with the transversely-polarized ferroelectric modes.

Strontium titanate ($SrTiO_3$) is among the most extensively studied phase transition materials, which displays intriguing properties such as ferroelectricity, superconductivity, and bears great potential for important technological applications such as thermoelectrics.[1-3] Enhancement of both Seebeck coefficient ($S$) and electrical conductivity ($\sigma$) and reduction of thermal conductivity ($\kappa$) are essential to escalate thermoelectric figure of merit at temperature $T$ ($ZT=S^2\sigma T/\kappa$).[4] Recently, the potential of $SrTiO_3$ as an environmentally-friendly oxide thermoelectric material was highlighted by a giant $S$ realized in $SrTi_{0.8}Nb_{0.2}O_3$ quantum well by electron confinement effect.[5] In fact, properly doped $SrTiO_3$ already exhibited a high power factor ($S^2\sigma$) comparable to that of $Bi_2Te_3$.[3] There, its intrinsic moderate lattice thermal conductivity ($\kappa_{lat} \approx \kappa$) is the bottleneck for further improvement of $ZT$.

In the target temperature range for thermoelectrics above its phase transition point ($T_c$=105 K), $SrTiO_3$ crystallizes in a cubic perovskite structure.[6] While there are several theoretical studies[6,7] devoted into understanding the electronic properties of perovskite $SrTiO_3$, its thermal transport properties have received less attention and remain an open question. Investigation of key thermal transport properties such as $\kappa_{lat}$ and phonon mean free path (MFP) so far has greatly relied on the empirical equations with either fitting or adjustable parameters,[8] and more accurate and systematic investigation on microscopic phonon transport is needed to gain further insight into possible strategies for $\kappa_{lat}$ reduction.

For accurate calculations of phonon transport, recently-established anharmonic lattice dynamics (ALD) method with interatomic force constants (IFCs) obtained from first principles has been demonstrated to be a powerful tool.[9] It has been successfully applied to various thermoelectric materials[10-13] to obtain $\kappa_{lat}$ and to gain deep understanding in microscopic mechanism of thermal transport. However, conventional methods to obtain

IFCs based on the perturbation around ground state are no longer valid for perovskite $SrTiO_3$ due to the structural instability. On the other hand, it has been recently shown that first-principles molecular dynamics (FPMD) method can readily include any orders of anharmonicity at finite temperature,[14] and has been employed together with lattice dynamics to successfully compute phonon dispersion for phase transition $CaSiO_3$.[15] Here, we applied similar method combining FPMD and ALD[13,14] to perovskite $SrTiO_3$ to investigate harmonic and anharmonic phonon transport properties.

In present work, we employed the *real-space displacement* method[16] to calculate IFCs, which are defined as Taylor expansion coefficients of atomic force ($F$) with respect to displacement ($u$) from equilibrium position:

$$F_i^{\alpha} = -\sum_{j,\beta} \Phi_{ij}^{\alpha\beta} u_j^{\beta} - \frac{1}{2!} \sum_{jk,\beta\gamma} \Psi_{ijk}^{\alpha\beta\gamma} u_j^{\beta} u_k^{\gamma} + \dots, \quad (1)$$

where $\Phi$ and $\Psi$ are harmonic and cubic anharmonic IFCs, respectively. The integer $i$, $j$, and $k$ represent atomic indices, and $\alpha$, $\beta$, and $\gamma$ are Cartesian components. As mentioned above, the conventional method calculates IFCs using $F$ and $u$ sampled at ground state, which would lead to imaginary phonon frequencies for perovskite $SrTiO_3$, i.e. thermodynamically unstable. We thus performed FPMD simulation to stabilize perovskite structure of $SrTiO_3$, to sample the $F$ and $u$ datasets.

The FPMD simulation was conducted for $4\times4\times4$ supercell containing 320 atoms using plane-wave basis method implemented in the *Quantum Espresso* package.[17] We adopted both local density approximation (LDA) and generalized gradient approximation (GGA) for exchange-correlation functional to study pseudopotential-dependent phonon properties. The parameterizations of Perdew-Wang and Perdew-Burke-Ernzerhof were selected for LDA and GGA, respectively. The cutoff energies for plane-wave expansion and charge density were set to 70(60) and 280(700) Ryd for LDA (GGA). Γ point was chosen for *k*-point sampling. The lattice constants for LDA and GGA were set to 3.81 and 3.93 Å, respectively, which were obtained by minimization of total energy.

In the FPMD simulation, the system was equilibrated at 300 K (NVT) using the velocity scaling method for ~0.25 ps with time step of 0.483 fs followed by ~0.5 ps-long constant energy (NVE) run for collecting atomic displacements and forces datasets. Temperature fluctuation during FPMD-NVE run is less than 5%, and root mean squared displacements of constituent elements are in agreements with experiment[18)] (Tab. 1). We then calculated harmonic and cubic IFCs using ALAMODE package.[13)] As for the range of IFCs, we included all allowed interactions in $4 \times 4 \times 4$ supercell for harmonic IFCs, and second nearest neighbors for cubic IFCs. Once harmonic IFCs are obtained, phonon dispersion relations can be calculated by solving dynamical matrix with given wavevectors. Polar effect is modeled by adding a nonanalytical term with mixed-space method.[19)] The Born effective charges of constituent elements and dielectric constant were calculated through macroscopic electric polarization using density function perturbation theory (Tab. 1).[17)]

Figure. 1 shows phonon dispersion relations of perovskite $SrTiO_3$ computed with LDA and GGA. Unlike the previous calculations based on ground-state perturbation[23,24)] that gave imaginary frequencies for anti-ferroelectric distortive (AFD) modes (R and M points) and ferroelectric (FE) mode (zone center Γ point), the current FPMD-based calculation with both LDA and GGA succeeded in reproducing non-imaginary frequencies of perovskite $SrTiO_3$. Phonon frequencies calculated with LDA appear higher (harder) through the entire Brillouin zone and those with GGA better reproduce the experiments.[25-27)] A significant pseudopotential dependence of FE mode around 2.5 THz is observed, which has been also found in previous calculations.[23,24)] The good agreements of AFD and FE modes with experiments at 300 K indicate the capability of current FPMD simulation to take finite temperature effect into account. Note that except for [100] direction, dispersions of longitudinal acoustic phonons (LA) are quickly flattened as wavevector goes away from zone center. By contrast to acoustic branches, some of the optical branches are quite dispersive, which implies that their contribution to $\kappa_{\mathrm{lat}}$ are not negligible.

In addition to phonon dispersion relation, to further validate the obtained harmonic IFCs, we calculated temperature-dependent volumetric heat capacity ($C_{\mathrm{V}}$) shown in Fig. 2. $C_{\mathrm{V}}$ calculated with LDA underestimates experiment at low temperatures due to hardening of phonon frequencies (Fig. 1). $C_{\mathrm{V}}$ calculated from phonon dispersion relation reasonably agrees with experiments[28-30] for the range of temperatures above $T_{\mathrm{c}}$ despite that the IFCs are calculated for 300 K. This is because frequency shifts of most phonons except for AFD and FE modes do not strongly depend on temperature (Fig. 1), which implies that it is reasonable to use harmonic IFCs obtained at 300 K for investigating phonon transport in the wide temperature range. We also extracted Debye temperatures from $C_{\mathrm{V}}$ and listed them in Tab.1.

To gain insight into the lattice anharmonicity, we next calculated mode-Grüneisen parameters of phonons with wavevector **q** and branch $j$ using the obtained cubic IFCs. Grüneisen parameter is a response of phonon frequency to volume change, and quantifies anharmonicity of crystal lattice[31]: $\gamma_j(\mathbf{q})=(-\partial \ln\omega_j(\mathbf{q})/\partial \ln V)$, where $V$ is crystal volume and $\omega_j(\mathbf{q})$ denotes frequency of ($\mathbf{q}$, $j$) phonon. Again, as shown in Fig. 3, significant pseudopotential dependence of FE mode manifests itself with large difference between the values of LDA ($\gamma_j(\mathbf{q})$ ~ 3.5) and GGA ($\gamma_j(\mathbf{q})$ ~ 26.8). Optical phonons along M-R symmetry line show relatively large values (-8.6 at maximum), highlighting intrinsic structural instability arising from such phonons as suggested by previous researches.[23,24] Similar divergence of FE mode have been seen in lead chalcogenides.[12,32] We looked into the calculated irreducible cubic IFCs in both GGA and LDA and found some of them closely associated with atomic motions for FE and AFD modes are considerably larger than the average value of cubic IFCs. Therefore, such anharmonic modes stems from the relevant motions governed by those large cubic IFCs, and results in low optical phonon frequencies (around 2.5 THz) observed in Fig. 1 despite relatively simple perovskite structure.

We also evaluated linear thermal expansion coefficient ($\alpha$) to validate obtained cubic IFCs, which is given by

$$\alpha = \frac{\gamma_{th} C_V}{3BV} . \tag{2}$$

Here $B$ represents bulk modulus, which is calculated from total energy calculation and equation of states[33] (Tab.1). $\gamma_{\mathrm{th}}$ is Grüneisen constant defined as an average of mode-Grüneisen parameters weighted by heat capacities of phonons in entire Brillouin zone:

$$\gamma_{th}(T) = \frac{1}{C_V(T)} \sum_{\mathbf{q},j} \gamma_j(\mathbf{q}) C_{V,j}(\mathbf{q},T) . \tag{3}$$

As a result, $\gamma_{\mathrm{th}}$ calculated with LDA and GGA at 300 K are 1.454 and 1.462, which are comparable with previous results.[34,35] Furthermore, calculated $\alpha$ are $0.608\times10^{-5}$ $K^{-1}$ (LDA) and $0.795\times10^{-5}$ $K^{-1}$ (GGA) at 300 K, which are consistent with experiments at the same temperature ($0.86$-$0.88\times10^{-5}$ $K^{-1}$).[36,37]

We finally calculate $\kappa_{\mathrm{lat}}$ based on phonon gas model:

$$\kappa_{lat} = \frac{1}{3V} \sum_{\mathbf{q},j} C_j(\mathbf{q}) v_{g,j}^2(\mathbf{q}) \tau_j(\mathbf{q}) , \tag{4}$$

where $v_{g,j}(\mathbf{q})$ and $\tau_j(\mathbf{q})$ are group velocity and relaxation time of ($\mathbf{q}$, $j$) phonon. The ALD calculation was performed on $20\times20\times20$ uniform $q$-meshes. Fig. 4 shows calculated temperature-dependent $\kappa_{\mathrm{lat}}$ together with the values from various experiments.[38-42] Note that thermal conductivity due to excited carriers is negligible up to 900 K since perovskite $SrTiO_3$ is a wide-gap semiconductor. As seen in Fig. 4, GGA well-reproduces the temperature dependence of $\kappa_{lat}$ at temperature higher than 300 K and LDA overestimates $\kappa_{lat}$ in the same temperature range because of smaller anharmonicity as indicated by Grüneisen parameters (Fig. 3). The relatively large deviation from experiments at low temperatures is attributed to sensitivity of $\kappa_{lat}$ to the possible impurity (e.g. oxygen atoms deficiency) scattering and finite-size effect particularly dominant in this temperature range. The MFP dependence of cumulative $\kappa_{lat}$ (an accumulation function of $\kappa_{lat}$)[43] shown in inset suggests phonons under 10 nm contributes considerably to heat conduction at room temperature. This indicates that very small structures would be required if one aims to reduce $\kappa_{lat}$ by nanostructuring.[44]

To further investigate mode-dependence heat conduction, we computed cumulative $\kappa_{lat}$ in frequency space at 300 K (Fig. 5). In low frequency regime, cumulative $\kappa_{lat}$ increases monotonically due to contribution from acoustic phonons. However, their contribution is quite limited and less than 30%. Surprisingly, optical phonons above 2.5 THz (especially transverse optical phonons, TO) contribute to over 60% of total $\kappa_{lat}$ and are the dominant heat carriers in perovskite $SrTiO_3$. To gain insights into the mechanism, we compared the mode-dependent $\kappa_{lat}$ along representative symmetry lines (Γ-X and Γ-R) calculated using LDA and GGA. As seen in Fig. 3, the calculations using LDA and GGA serve as the case studies without and with the divergence in Grüneisen number (i.e. anharmonicity) at FE mode (TO phonons at Γ). As summarized in Tab. 2, the comparison suggests that the presence of the divergence significantly reduces $\kappa_{lat}$ of both transverse (TA) and longitudinal (LA) acoustic phonons. The strong coupling between TA and TO phonons has been previously suggested in relaxor ferroelectrics.[45] On the other hand, the strong LA-TO coupling is consistent with what has been seen for incipient ferroelectric lead telluride.[12,46]

In summary, we have applied first principles phonon transport analysis combining FPMD

and ALD to perovskite $SrTiO_3$, a phase transition material which is stable only above 105 K. By incorporating finite temperature effect into harmonic IFCs, we successfully computed phonon dispersion relations that exhibit non-imaginary frequencies and match with experiments. The calculations further reproduced temperature dependent lattice thermal conductivity well and disclosed an exceptionally large heat conduction contribution from optical phonons. The mode- and pseudopotential-dependent analysis suggest that the small contribution of acoustic phonons results from coexistence of TA-TO and LA-TO coupling arising from strongly anharmonic FE mode in perovskite $SrTiO_3$. In addition, demonstrated applicability of current approach to the phase transition material displays a great potential to extend first-principles-based phonon transport analysis to wide range of materials, where exciting phenomena remain to be uncovered.

## Acknowledgments

This work is supported partially by KAKENHI (Grand numbers: 26709009, 26630061, and 15K17982), Thermal and Electrical Energy Technology Foundation, and Doctoral Student Special Incentives Program at Graduate School of Engineering, The University of Tokyo (SEUT RA). The authors are grateful for fruitful discussions with Dr. Terumasa Tadano.

## References

1) M. L. Cohen, Phys. Rev. **134**, A511 (1964).
2) J. G. Bednoiz and K. A. Muller, Phys. Rev. Lett. **52**, 2289 (1984)
3) T. Okuda, K. Nakanishi, S. Miyasaka, and Y. Tokura, Phys. Rev. B **63**, 113104 (2001).
4) H. J. Goldsmid, *Introduction to Thermoelectricity* (Springer, Heidelberg, 2009).
5) H. Ohta, S. Kim, Y. Mune, T. Mizoguchi, K. Nomura, S. Ohta, T. Nomura, Y. Nakanishi, M. Hirano, H. Hosono, and K. Koumoto, Nat. Mater. **6**, 129 (2007).
6) W. Luo, W. Duan, S. G. Louie, and M. L. Cohen, Phys. Rev. B **70**, 214109 (2004).
7) M. Marques, L. K. Teles, V. Anjos, L. M. R. Scolfaro, J. R. Leite, V. N. Freire, G. A. Farias, and E. F. da Silva, Jr., Appl. Phys. Lett. **82**, 3074 (2003).
8) Y. Wang, K. Fujinami, R. Zhang, C. Wan, N. Wang, Y. Ba, and K. Koumoto, Appl. Phys. Express **3**, 031101 (2010).
9) D. A. Broido, M. Malorny, G. Birner, N. Mingo, and D. A. Stewart, Appl. Phys. Lett. **91**, 231922 (2007).
10) K. Esfarjani, G. Chen, and H. T. Stokes, Phys. Rev. B **84**, 085204 (2011).
11) J. Shiomi, K. Esfarjani, and G. Chen, Phys. Rev. B **84**, 104302 (2011).
12) T. Shiga, J. Shiomi, J. Ma, O. Delaire, T. Radzynski, A. Lusakowski, K. Esfarjani, and G. Chen, Phys. Rev. B **85**, 155203 (2012).
13) T. Tadano, Y. Gohda, and S. Tsuneyuki, J. Phys.: Condens. Matter **26**, 225402 (2014).
14) O. Hellman and D. A. Broido, Phys. Rev. B **90**, 134309 (2014).
15) T. Sun, D.-B. Zhang, and R. M. Wentzcovitch, Phys. Rev. B **89**, 094109 (2014).
16) K. Esfarjani and H. T. Stokes, Phys. Rev. B **77**, 144112 (2008).
17) G. Paolo, B. Stefano, B. Nicola, et al., J. Phys.: Condens. Matter **21**, 395502 (2009).
18) Y. A. Abramov, V. G. Tsirelson, V. E. Zavodnik, S. A. Ivanov, and Brown I. D., Acta Crystallogr. Sect. B **51**, 942 (1995).
19) Y. Wang, J. J. Wang, W. Y. Wang, Z. G. Mei, S. L. Shang, L. Q. Chen, and Z. K. Liu, J. Phys.: Condens. Matter **22**, 202201 (2010).
20) S. Piskunov, E. Heifets, R. I. Eglitis, and G. Borstel, Comput. Mater. Sci. **29**, 165 (2004).
21) G. Burns, Solid State Commun. **35**, 811 (1980).
22) A. Duran, F. Morales, L. Fuentes, and J. M. Siqueiros, J. Phys.: Condens. Matter **20**, 085219 (2008).
23) C. Lasota, C.-Z. Wang, R. Yu, and H. Krakauer, Ferroelectrics **194**, 109 (1997).
24) R. A. Evarestov, E. Blokhin, D. Gryaznov, E. A. Kotomin, and J. Maier, Phys. Rev. B **83**, 134108 (2011).
25) R. A. Cowley, Phys. Rev. **134**, A981 (1964).
26) R. A. Cowley, W. J. L. Buyers, and G. Dolling, Solid State Commun. **7**, 181 (1969).
27) W. G. Stirling, J. Phys. C: Solid State Phys. **5**, 2711 (1972).
28) J. P. Coughlin and R. L. Orr, J. Am. Chem. Soc. **75**, 530 (1953).
29) D. de Ligny and P. Richet, Phys. Rev. B **53**, 3013 (1996).
30) S. S. Todd and R. E. Lorenson, J. Am. Chem. Soc. **74**, 2043 (1952).
31) G. P. Srivastava, *The Physics of Phonons* (Taylor & Francis, 1990).
32) Y. Zhang, X. Ke, C. Chen, J. Yang, and P. R. C. Kent, Phys. Rev. B **80**, 024304 (2009).
33) F. Birch, J. Appl. Phys. **9**, 279 (1938).
34) H. Ledbetter, Phys. C: Superconductivity **159**, 488 (1989).
35) A. G. Beattie and G. A. Samara, J. Appl. Phys. **42**, 2376 (1971).
36) K. Itoh, K. Ochiai, H. Kawaguchi, C. Moriyoshi, and E. Nakamura, Ferroelectrics **159**, 85 (1994).

37) K. Munakata and A. Okazaki, Acta Crystallogr. Sect. A **60**, 33 (2004).
38) M. T. Buscaglia, F. Maglia, U. Anselmi-Tamburini, D. Marré, I. Pallecchi, A. Ianculescu, G. Canu, M. Viviani, M. Fabrizio, and V. Buscaglia, J. Eur. Ceram. Soc. **34**, 307 (2014).
39) Y. Suemune, J. Phys. Soc. Jpn. **20**, 174 (1965).
40) H. Muta, K. Kurosaki, and S. Yamanaka, J. Alloys Compd. **392**, 306 (2005).
41) M. Ito and T. Matsuda, J. Alloys Compd. **477**, 473 (2009).
42) C. Yu, M. L. Scullin, M. Huijben, R. Ramesh, and A. Majumdar, Appl. Phys. Lett. **92**, 191911 (2008).
43) C. Dames and G. Chen, *Thermoelectrics Handbook: Macro to Nano* (CRC Press, Boca Raton, 2005).
44) D. Aketo, T. Shiga, and J. Shiomi, Appl. Phys. Lett. **105**, 131901 (2014).
45) J. Hlinka, S. Kamba, J. Petzelt, J. Kulda, C. A. Randall, and S. J. Zhang, Phys. Rev. Lett. **91**, 107602 (2003).
46) J. An, A. Subedi, and D. J. Singh, Solid State Commun. **148**, 417 (2008).

## Figure Captions

**Fig. 1.** Phonon dispersion relations of $SrTiO_3$ computed with (a) LDA and (b) GGA along high symmetry lines. The marks denote inelastic neutron scattering experiments at 90 K (blue open circles), 297 and 300 K (red open circles). [25-27]

**Fig. 2.** Temperature dependence of constant volume heat capacity ($C_V$) calculated from phonon dispersion relations in Fig. 1. The marks denote experimental data. [28-30]

**Fig. 3.** Mode-Grüneisen parameters using anharmonic cubic IFCs along high symmetry lines.

**Fig. 4.** Temperature dependence of $\kappa_{lat}$ calculated with LDA and GGA. The marks denote experimental data.[38-42] The inset figure shows MFP dependence of cumulative lattice thermal conductivity ($\kappa_{cum}$) normalized by overall lattice thermal conductivity ($\kappa_{bulk}$) at 300 K.

**Fig. 5.** Frequency-dependent cumulative lattice thermal conductivity ($\kappa_{cum}$) normalized by overall lattice thermal conductivity ($\kappa_{bulk}$) at 300 K with (a) LDA and (b) GGA. The right y axis denotes intensity of phonon density of states (DOS).

## Table Captions

**Tab 1.** Several calculated quantities of perovskite $SrTiO_3$ and experimental data.

**Tab 2.** Branches decomposed $\kappa_{lat}$ along Γ-X and Γ-R direction with LDA and GGA.

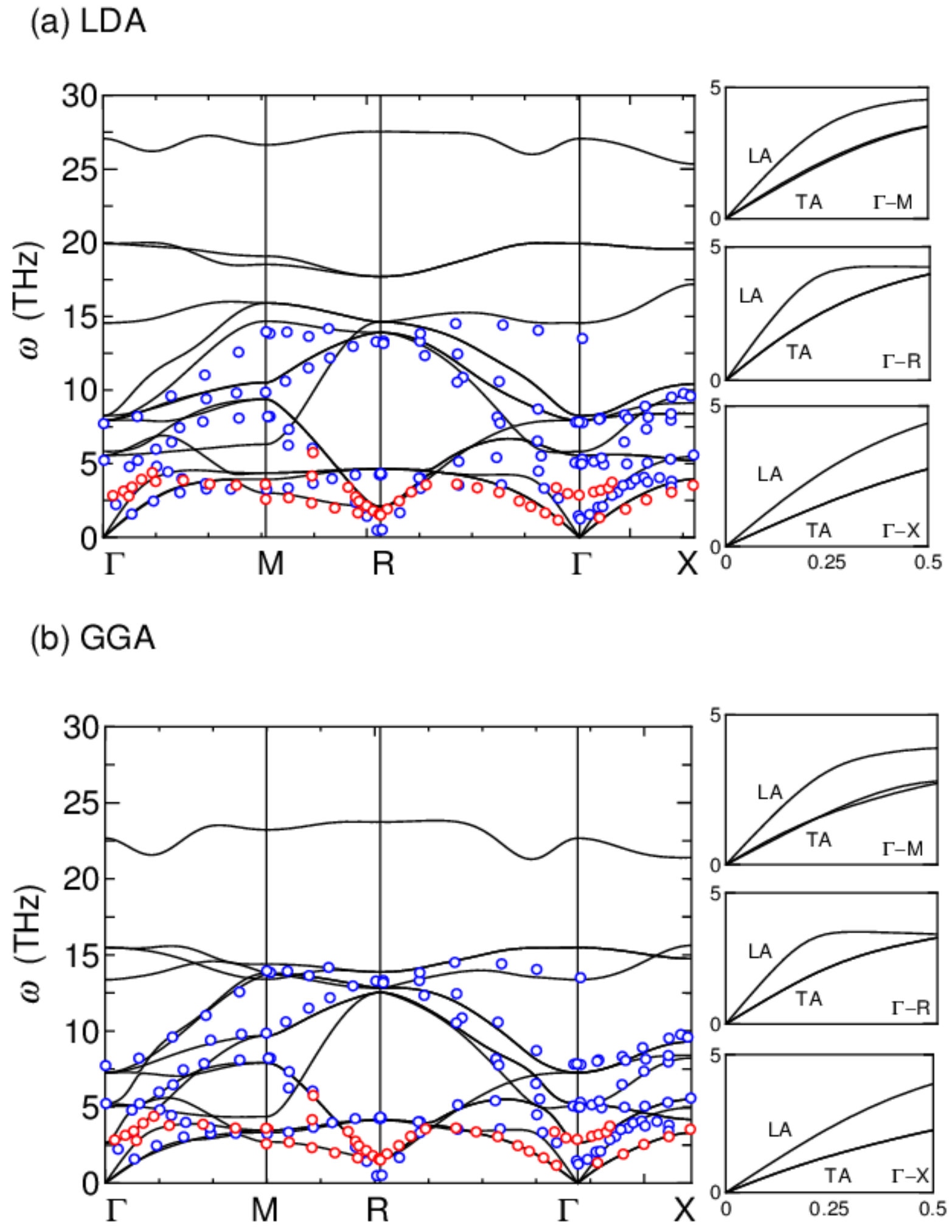

Fig. 1

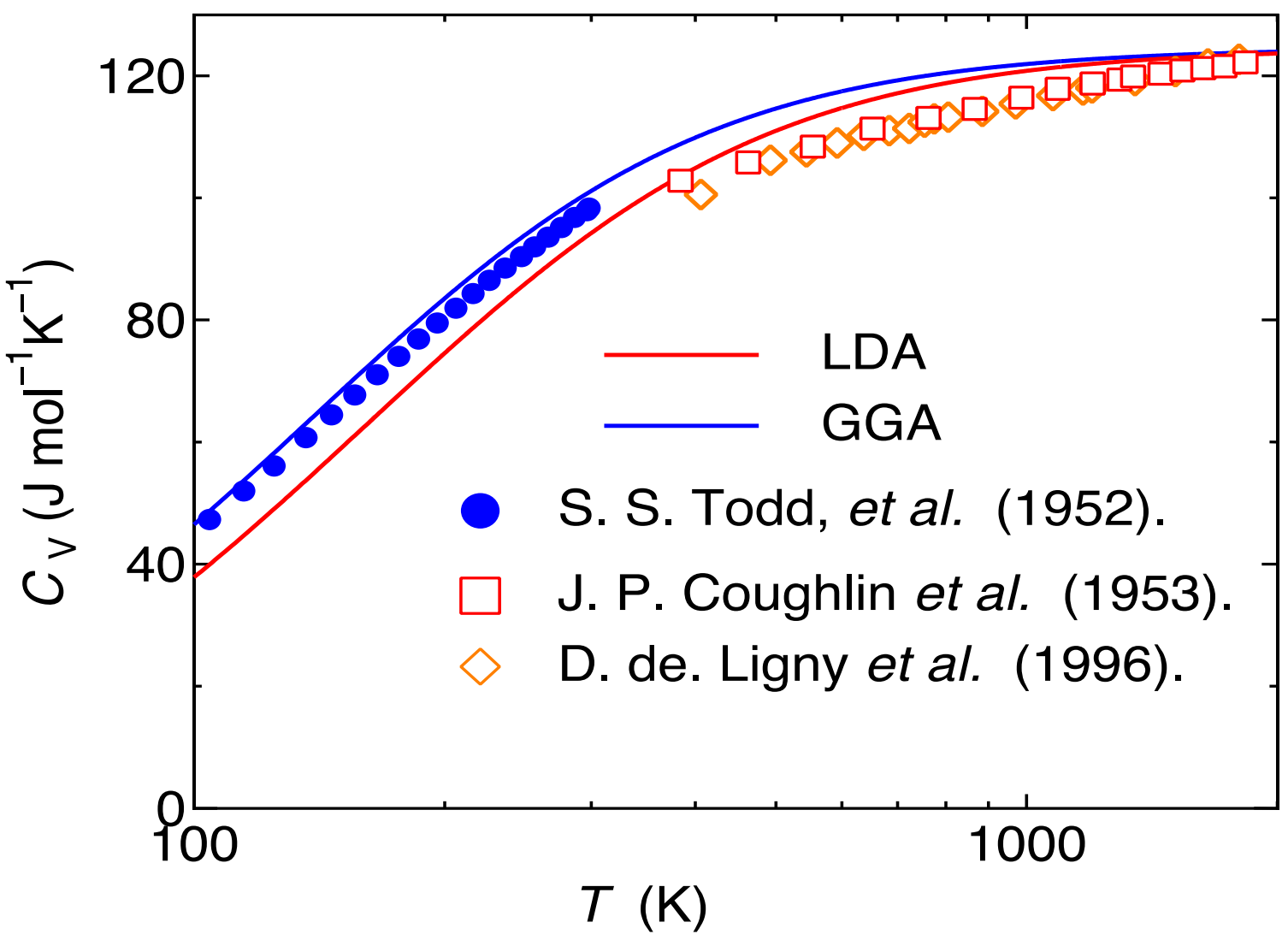

120
80
40
0
100
1000
$C_V$ (J mol$^{-1}$K$^{-1}$)
$T$ (K)
LDA
GGA
S. S. Todd, *et al.* (1952).
J. P. Coughlin *et al.* (1953).
D. de. Ligny *et al.* (1996).

Fig. 2

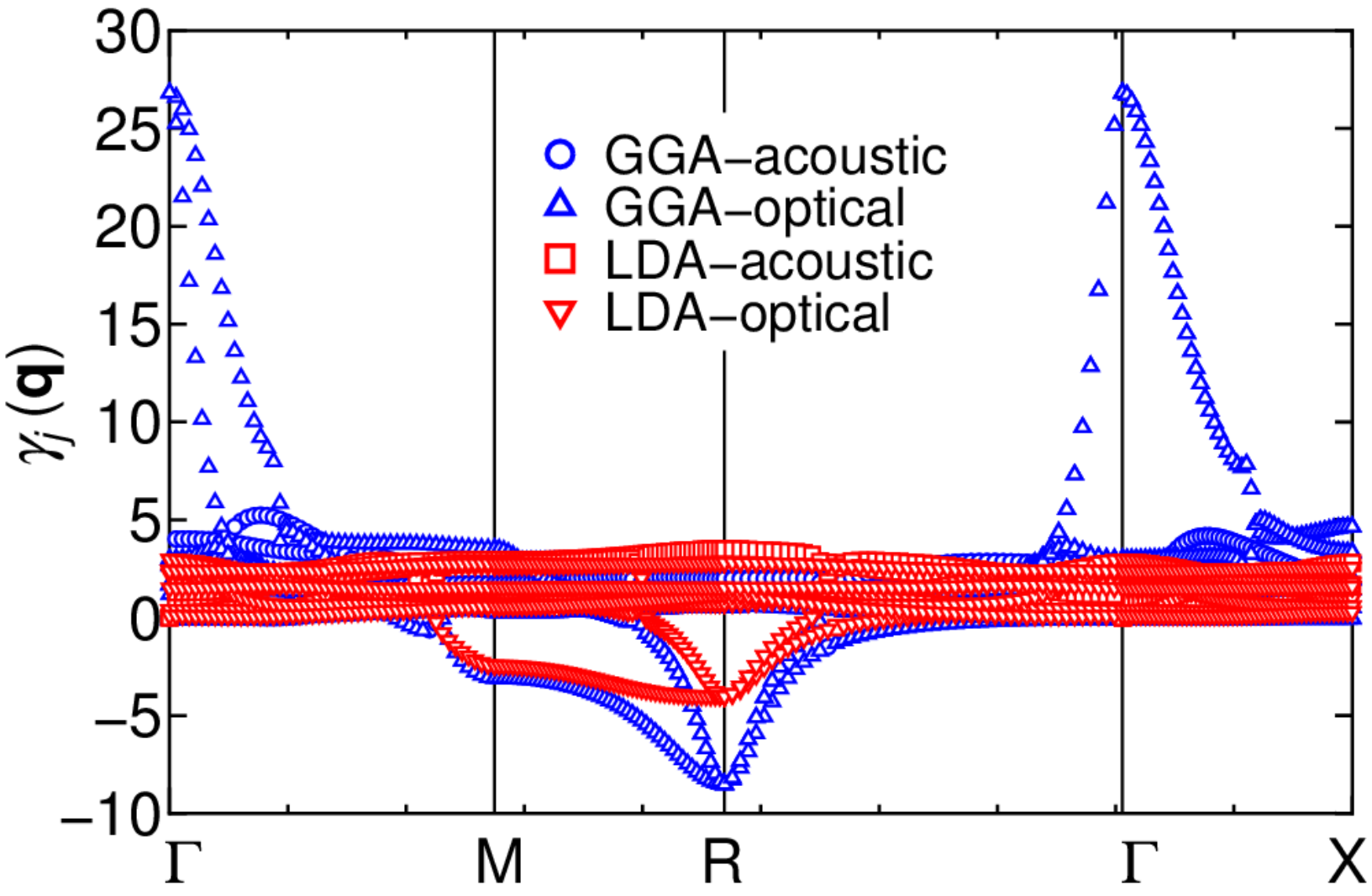

Fig. 3

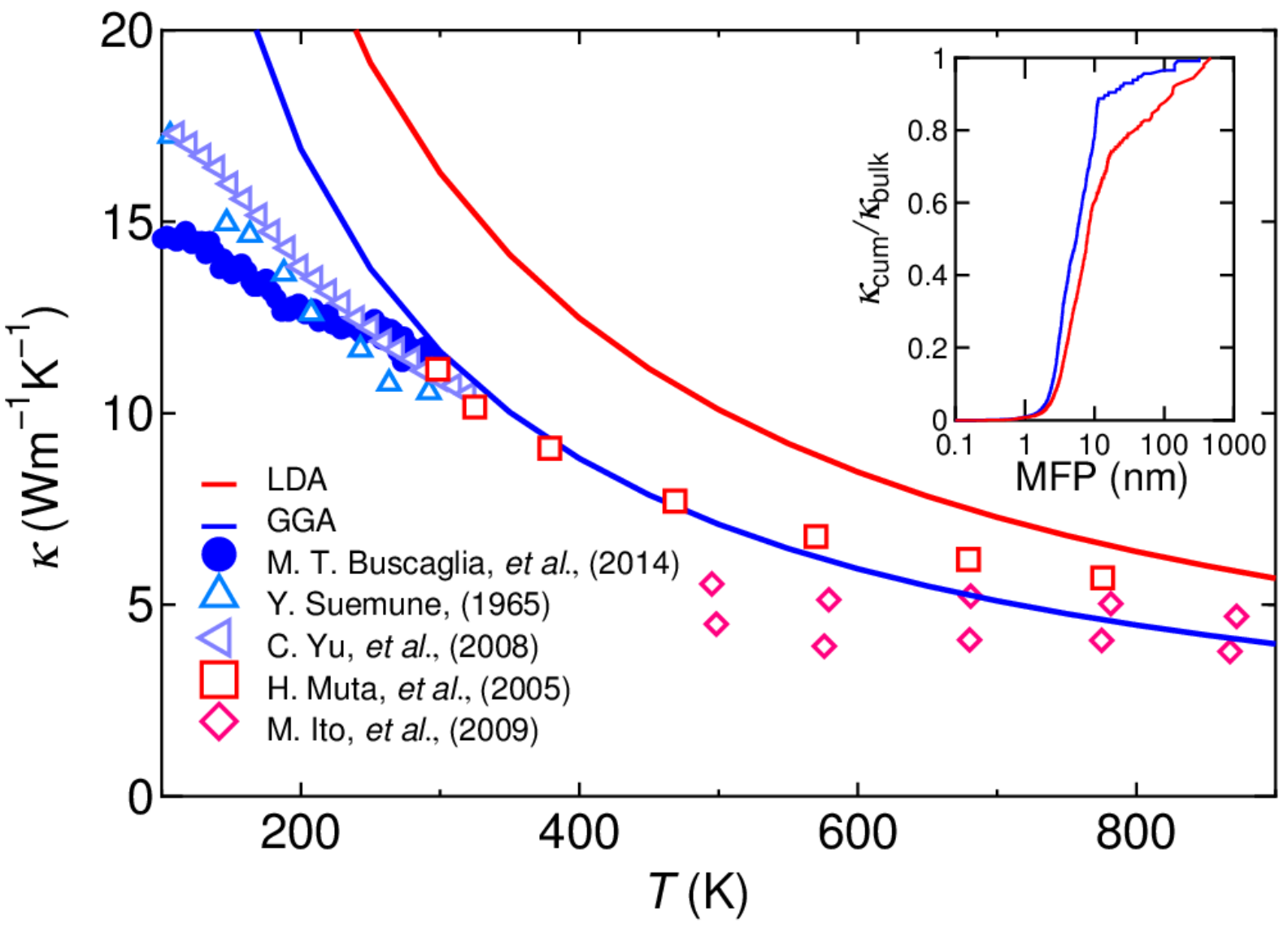

κ (Wm−1K−1)
T (K)
LDA
GGA
M. T. Buscaglia, et al., (2014)
Y. Suemune, (1965)
C. Yu, et al., (2008)
H. Muta, et al., (2005)
M. Ito, et al., (2009)
κcum/κbulk
MFP (nm)

Fig. 4

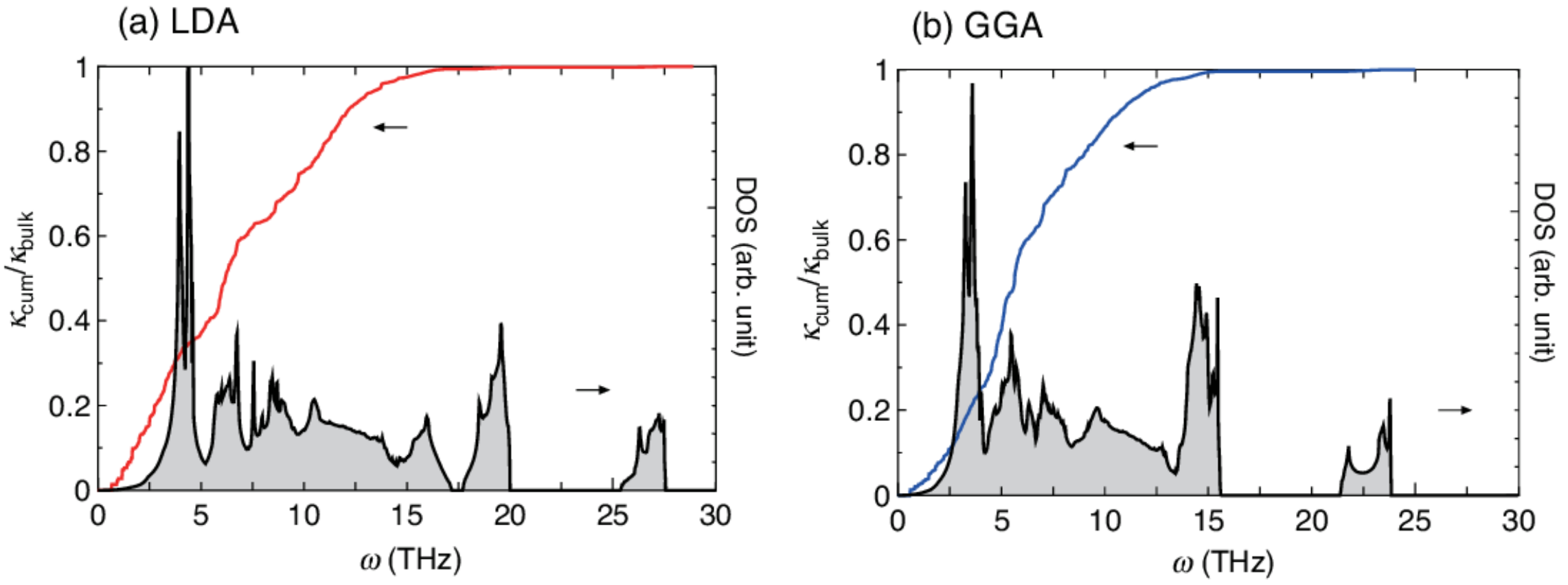

Fig. 5

Tab. 1

| | |
|---|---|
| Root mean squared displacement, $u$ (Å) | |
| (LDA)<br>$u$(Sr)=0.072, $u$(O)=0.106, $u$(Ti)=0.046<br>(GGA)<br>$u$(Sr)=0.091, $u$(O)=0.104, $u$(Ti)=0.068 | (Exp.)[18]<br>$u$(Sr)=0.089,<br>$u$(O)=0.105,0.069<br>$u$(Ti)=0.75 |
| Born effective charge, $Z^*$ ($e$) | |
| (LDA)<br>$Z^*$(Sr)=2.27, $Z^*$(Ti)=6.75,<br>$Z^*(O)_{\parallel}$=-5.29, $Z^*(O)_{\perp}$=-1.86<br>(GGA)<br>$Z^*$(Sr)=2.43, $Z^*$(Ti)=7.4,<br>$Z^*(O)_{\parallel}$=-5.87, $Z^*(O)_{\perp}$=-2.03 | (Calc.)[23]<br>$Z^*$(Sr)=2.55<br>$Z^*$(Ti)=7.56<br>$Z^*(O)_{\parallel}$=-5.92<br>$Z^*(O)_{\perp}$=-2.12 |
| Dielectric constant, $\varepsilon_{\infty}$ | |
| (LDA) 5.3, (GGA) 6.3 | (Calc.)[23] 6.63 |
| Bulk modulus, $B$ (GPa) | |
| 225.1 (LDA), 169.4 (GGA) | (Calc.) 214[20], 169[20] |
| Debye Temperature (K) | |
| (LDA) 731.5, (GGA) 627.4 | (Exp.)633[21], 640[22] |

Tab. 2

| $\kappa_{lat}$ ($10^{-2}$ $Wm^{-1}K^{-1}$) | | | | |
|---|---|---|---|---|
| | Γ-X | | Γ-R | |
| | LDA | GGA | LDA | GGA |
| TA | 5.98 | 3.44 | 1.45 | 0.48 |
| LA | 5.93 | 1.23 | 2.55 | 1.23 |